\documentclass[aip, jcp, amsmath, amssymb, twocolumn, preprint]{revtex4-1}

\usepackage[utf8x]{inputenc}
\usepackage{amsmath}
\usepackage{amsfonts}
\usepackage{graphicx}
\usepackage{multirow}
\usepackage{color}
\usepackage{txfonts}
\usepackage{textcomp}
\usepackage{float}
\usepackage{subfigure}

\newcommand{\de}{\mathrm{d}}
\newcommand{\e}{\mathrm{e}}

\begin{document}

\title{A one-dimensional model with water-like anomalies and two phase transitions}
\author{Lotta Heckmann}
\thanks{lotta@fkp.tu-darmstadt.de}
\author{Barbara Drossel}
\affiliation{Institut f\"ur Festk\"orperphysik, Technische Universit\"at Darmstadt, Hochschulstr. 6, 64289 Darmstadt, Germany}

\begin{abstract}

We investigate a one-dimensional model that shows several properties of water. The model combines the long-range attraction of the van der Waals model with the  nearest-neighbor interaction potential by Ben-Naim, which is a step potential that includes a hard core and a potential well. Starting from the analytical expression for the partition function, we determine numerically the Gibbs energy and other thermodynamic quantities. The model shows two phase transitions, which can be interpreted as the liquid-gas transition and a transition between a high-density and a low-density liquid. At zero temperature, the low-density liquid goes into the crystalline phase. Furthermore, we find several anomalies that are considered characteristic for water. We explore a wide range of pressure and temperature values and the dependence of the results on the depth and width of the potential well.
 \end{abstract}

\maketitle

\section{Introduction}
Due to its omnipresence on earth and its importance for life, understanding the behavior of water and water mixtures is of utmost importance. Although the anomalous properties of water have been subject to investigation for a long time, the reasons and underlying principles leading to the outstanding characteristics of water are not yet completely understood \cite{5_2007_stanley, nezbeda11}. Recent effort has focused on the properties at low temperatures and high pressures, where many different ice phases were found, and where a liquid-liquid phase transition between a high-density phase and a low-density phase was postulated 20 years ago \cite{poole92}. Despite the fact that the idea of characterizing liquid water by a high-density and a low-density liquid is old and has often been discussed \cite{bennaim1974}, this so-called LDL-HDL transition, which implies that water has a second critical point, is still controversial. Several alternative scenarios have been suggested, all of which can be described within a simple cooperative hydrogen-bond model by varying the model parameters \cite{stokely10}. Nevertheless, a LDL-HDL transition is a generic scenario for core-softened potentials \cite{hemmer70,stell72,scala00}, and the experimentally observed  dynamical crossover between fragile and strong behavior in supercooled water can be interpreted as a signature of the Widom line \cite{xu05}.

Theoretical studies of water range from being as precise as possible to being as simple as possible. One older example for a description of the behaviour of water that allows for the calculation (and thus prediction) of properties with large precision is an equation of state for water with 58 parameters \cite{saul1989}. With such complex descriptions, however, no mechanisms can be identified, and the underlying principles leading to the characteristic behaviour of water remain unrevealed.  Nowadays, the dominant theoretical approaches to water are molecular dynamics simulations and Monte Carlo simulations, which have become possible due to increasing computing power, allowing for a microscopic modelling of water starting from the full quantum mechanical description of its molecules, or from  simpler, classical molecular models with few parameters. These models were able to reproduce many properties of water \cite{1_2002_guillot}.

However, critical voices \cite{brodky96} argue that computer simulations of water do not follow the basic guidelines of theoretical investigations. Though promising in many respects, the systems consist of hundreds to thousands particles interacting via semi-empirical interaction potentials, for which the interpretation and identification of mechanisms remains difficult. A need has been identified to find even simpler models capturing relevant features of water in order to understand which ingredients are necessary and indispensable for the occurrence of anomalies or other features of real water. 

An important step in the understanding of water was the insight that neither the bent shape of the molecule nor the directedness of hydrogen bonds are necessary for many of the observed features of water. In fact, a soft core in a radially symmetric interaction potential is sufficient to  allow for a discontinuous change in the preferred distance between molecules as the pressure is varied, leading to various water anomalies \cite{vilaseca2011}. The idea of modelling water with an effective soft-core potential is experimentally supported by potentials derived from the experimental O-O radial distribution function \cite{headgordon93}. 

A further simplification is implemented by  the Jagla potential \cite{jagla98}, which has no attractive part at all, but a ramp that is a simple representation of a soft core. Even simpler are repulsive step potentials which were introduced in the 1970's \cite{young77} and are still investigated until today \cite{stishov02,fomin08, gribova09, fomin11}, although there are results showing that a single temperature- and pressure-independent potential may not be sufficient to capture all properties of water \cite{chaimovich09}.

Even one-dimensional water models have been used to describe water anomalies. 
A lattice-gas model, where distance 2 between nearest neighbors is associated with a stronger binding energy than distance 1, shows a zero-temperature critical point and thermodynamic and dynamic anomalies similar to water\cite{barbosa11}. A rather complex lattice model with two different repulsion scales and a mean field attraction leads to a rather rich phase diagram \cite{hoye2010}. 
Continuous one-dimensional models with a short-range soft-core potential produce various water anomalies \cite{sadrlahijany1999}, which occur also in the two-dimensional version \cite{scala2001}. A model with two wells, which was first published in 1992 \cite{bennaim1992}, was argued to provide an explanation for the density anomaly \cite{cho1996}. The simplest one-dimensional model that produces water anomalies was introduced by Bell \cite{bell1969} and re-discovered  by Ben-Naim \cite{bennaim08I}. It contains a step potential with a single well and shows among other properties a density anomaly and a minimum in the isothermal compressibility.   A drawback of such one-dimensional models with short-range interactions is, however, that they cannot show a phase transition at temperatures larger than zero \cite{vanhove50}.
Several general reviews about simple and simplified water models have been written \cite{dill2005, nezbeda11}, and a very good introduction into water models can be found in the book by Ben-Naim \cite{bennaimbook}.

In addition to these newer models for water, there exists also the phenomenological mean field approach of van der Waals, introduced in his PhD thesis in 1873, that has since then been discussed extensively. The properties of a one-dimensional model in the van der Waals limit, where the range of the force goes to infinity while its strength goes to zero, were discussed in the 1960's by several authors \cite{kac63, uhlenbeck1963, hemmer64}. Ideas for adjusting this model to give a better description of water were also discussed by Heidemann and Prausnitz \cite{heidemann76}, where a van der Waals model for fluids with associating molecules is introduced. It was noted, however, the van der Waals model generally fails for water because there are strong directed interactions in water that can not be modeled with a mean field attraction \cite{chandler1983}.

In this article, we will study a one-dimensional model that combines the features of the Ben-Naim model and the van der Waals model. Combining a simple water model with a long-range attraction was first done by E.A. Jagla \cite{jagla99}. He found that by adding a long-range attraction similar to the van der Waals gas to his above-mentioned three-dimensional model, the water-like properties are maintained and an additional phase transition can occur. Adding a long-range attraction to a one-dimensional model, as will be done in this paper, leads to the probably simplest possible one-dimensional model that shows two phase transitions and various water anomalies. The mean-field term turns the zero-temperature phase transition of the Ben-Naim model, which is due to the potential well, into a finite-temperature phase transition between two different liquid phases. Furthermore, the mean-field term introduces the liquid-vapor phase transition into the one-dimensional model.

\section{Model}
Our model is a combination of the one-dimensional model by Ben-Naim and the van der Waals model. We first describe these two ingredients of the model separately before we present the combined model.
\subsection{Ben-Naim model}
\begin{figure}
\includegraphics[width=0.45\textwidth]{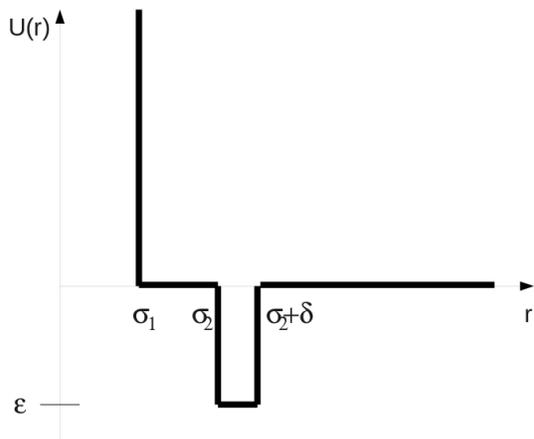}
  \caption{The short-range interaction potential for a 1D model of water suggested by Ben-Naim \cite{bennaim08I}.}
    \label{fig:BN_pot}
\end{figure}

The Ben-Naim model was first analyzed by Bell in 1969 \cite{bell1969} and discussed later with respect to water by Ben-Naim \cite{bennaim08I}. This model describes a one-dimensional system of $N$ particles interacting via a short-range potential. The potential sketched in figure \ref{fig:BN_pot} consists of a hard-core repulsion (corresponding to the excluded volume of a particle) and a minimum that mimics the effect of hydrogen bonding.
 The corresponding partition function can be calculated analytically, leading to the specific Gibbs energy
 \begin{align}
\label{eq:gBN}
 &g_{BN}(T,p) = \\ \nonumber
 & \frac{1}{\beta}  \ln \frac{\beta p\lambda}{\mathrm{e}^{-\beta p \sigma_1} + \mathrm{e}^{-\beta p \sigma_2}(\mathrm{e}^{-\beta \epsilon} -1) (1-\mathrm{e}^{-\beta p \delta}) }
\end{align}
 and to the specific volume 
\begin{align}
v_{BN}(T,p) &= \frac{\partial g}{\partial p}|_T  
\end{align}
with $\beta=\frac{1}{T}$ and the thermal wavelength $\lambda={h}/{\sqrt{2\pi m k_B T}}$. 
This model exhibits a density anomaly and a minimum of the isothermal compressibility, in agreement with two characteristic features of real water.
Additionally, for low temperatures a steep transition from a lower density to a higher density is observed with increasing pressure. At $T=0$, this transition becomes a real first-order phase transition, with all particles having distance $\sigma_1$ for $p(\sigma_2-\sigma_1)>|\epsilon|$ and distance $\sigma_2$ otherwise. This phase transition can be calculated by determining the phase that minimizes $G=E-TS+pV$  at $T=0$ and for the considered value of the pressure $p$.

\subsection{Van der Waals model}
The van der Waals gas is described by the thermal equation of state
\begin{align}
\left(p + \frac{a}{v^2} \right)(v-b)&= k_B T,
\end{align}
where $a$ and $b$ are parameters of the considered gas and $v$ is the specific volume $v=V/N$. For our model, we have $b=\sigma_1$. 
The pure van der Waals model leads to the well-known phase diagram of the van der Waals gas with a first order phase transition line between a fluid and a gas terminating in a critical point, which is situated at $T_c = 8a/27k_B b$, $p_c=a/27 b^ 2$ and $v_c =3b$.
If $a\neq 0$ and $b\neq 0$, the thermal equation of state is equivalent to
\begin{align}
\left(\tilde{p} + \frac{3}{\tilde{v}^2} \right)(3\tilde{v}-1)&= 8 \tilde{T}
\end{align}
with the reduced variables $\tilde{v} = v/v_c  ~ , ~ \tilde{T} = T/T_c ~ , ~\tilde{p} = p/p_c$ chosen such that the critical point is at $\tilde{v}=\tilde{p}=\tilde{T}=1$ generically \cite{stanley_book_1971}. 
Although this model has a liquid-gas phase transition, no other features of water are matched.
\subsection{The combined Ben-Naim--van der Waals model}
We now combine the two ingredients and investigate the resulting model, which we call the Ben-Naim--van der Waals model. Below, we will use the index BNJ for this model, because Jagla first introduced a van der Waals term into a water model.
The short-range interaction potential has the same form as the potential used by Ben-Naim (see figure \ref{fig:BN_pot}).
Additionally, we introduce a long-range interaction analogously to the van der Waals model via the substitution $p \rightarrow (p+a/v^2)$ in the above expression for the specific volume of the Ben-Naim model \cite{hemmer1976,jagla98,truskett2002_JCP}. The reduction of pressure by $ a /v^2$ follows directly from a decrease of the internal energy by $ a/ v$ due to the long-range attraction. We therefore obtain
\begin{align}
v(T,p) &\equiv v_{BNJ}(T,p) = v_{BN}(T,p+\frac{a}{v^2})\\
g(T,p) &\equiv g_{BNJ}(T,p) = g_{BN}(T,p+\frac{a}{v^2}) - \frac{2a}{v}.
\label{eq:gBNJ}
\end{align}
A more detailed derivation of expression (\ref{eq:gBNJ}) is given in Appendix B in the publication by Truskett and Dill \cite{truskett2002}.
The implicit equation for the specific volume can be solved numerically.
We use again the reduced variables $(\tilde{p}, \tilde{T},\tilde{v} )$ and introduce further the dimensionless variables 
$\tilde{\sigma_1} = \sigma_1/v_c  ~ , ~  \tilde{\sigma_2} = {\sigma_2}/{v_c} ~~ , ~~ \tilde{\delta} = {\delta}/{v_c} ~~ , ~~ \tilde{\epsilon} = {\epsilon}/{k_B T_c}  ~~ , ~~ \tilde{g} = {g}/{k_B T_c}$,
which are chosen such that lengths are measured in units of $v_c$ and that for $\tilde{\epsilon}=0$, the model reduces to the pure van der Waals system with a critical point at $\tilde{v}=\tilde{T}=\tilde{p}=1$.
Note that since $T_c \propto a$, we have $\tilde{\epsilon} \propto {\epsilon}/{a}$ and the reduced variable $\tilde{\epsilon}$ measures thus the ratio between the depth of the short-range potential well $\epsilon$ and the global attraction strength $a$.\\
This model has now three remaining free parameters
$\tilde{\epsilon}$, $\tilde{\sigma_2}$ and $\tilde{\delta}$. Note that since the hard core distance is $b=\sigma_1$, we have $\tilde{\sigma_1}={1}/{3}$, and this parameter has thus been eliminated.
From now on, we will omit the tilde, and we will set $\tilde{x}\equiv x$ for the variables $x=T,p,v,\sigma_1, \sigma_2, \delta, \epsilon, g$, which is equivalent to using dimensionless variables as defined above. This choice is not possible for the pure Ben-Naim model, where other (arbitrary) units are used.\\
 The expression for the specific volume $v(T,p)$ can be evaluated numerically in order to obtain a density profile. Whereever there is more than one solution for $ v$, one must determine the solution that minimizes the Gibbs energy in order to obtain a phase diagram. We also evaluated the isothermal compressibility and the isobaric heat capacity 
\begin{align}
\kappa_T &= - \frac{1}{v} \frac{\partial v}{\partial p}|_T\\
c_p &= T \frac{\partial s}{\partial T}|_p = -T \frac{\partial^2 g}{\partial T^2}|_p.
\label{eq:ktcp}
\end{align}
The first can be extracted from $v(T,p)$, and the second follows from equation (\ref{eq:gBNJ}).
Additionally, the probability density of the distance between nearest neighbours can be calculated. It is identical to the weight occuring in the expression for the partition function and is given by 
 \begin{align}
 q(r)&=\alpha \e^{-\frac{U(r)}{T}} \e^{-\frac{3(p+3/v^ 2)}{8T} r}
 \label{eq:qofr}
 \end{align}
where $\alpha$ is a normalization constant ensuring that $\int_0^\infty q(r) \de r =1$.

\section{Results}
If not noted otherwise, the parameters of the model are ${\sigma_2}=2/3$ and ${\delta}=1/10$.
These values are chosen such that their ratio matches that of the parameters used by Ben-Naim \cite{bennaim08I} (while we have to set ${\sigma_1}=1/3$).
The parameter $\epsilon$ is varied from $0$ down to $-10$.

\subsection{Density profiles and phase diagrams}
\begin{figure*}
\includegraphics[width=0.92\textwidth]{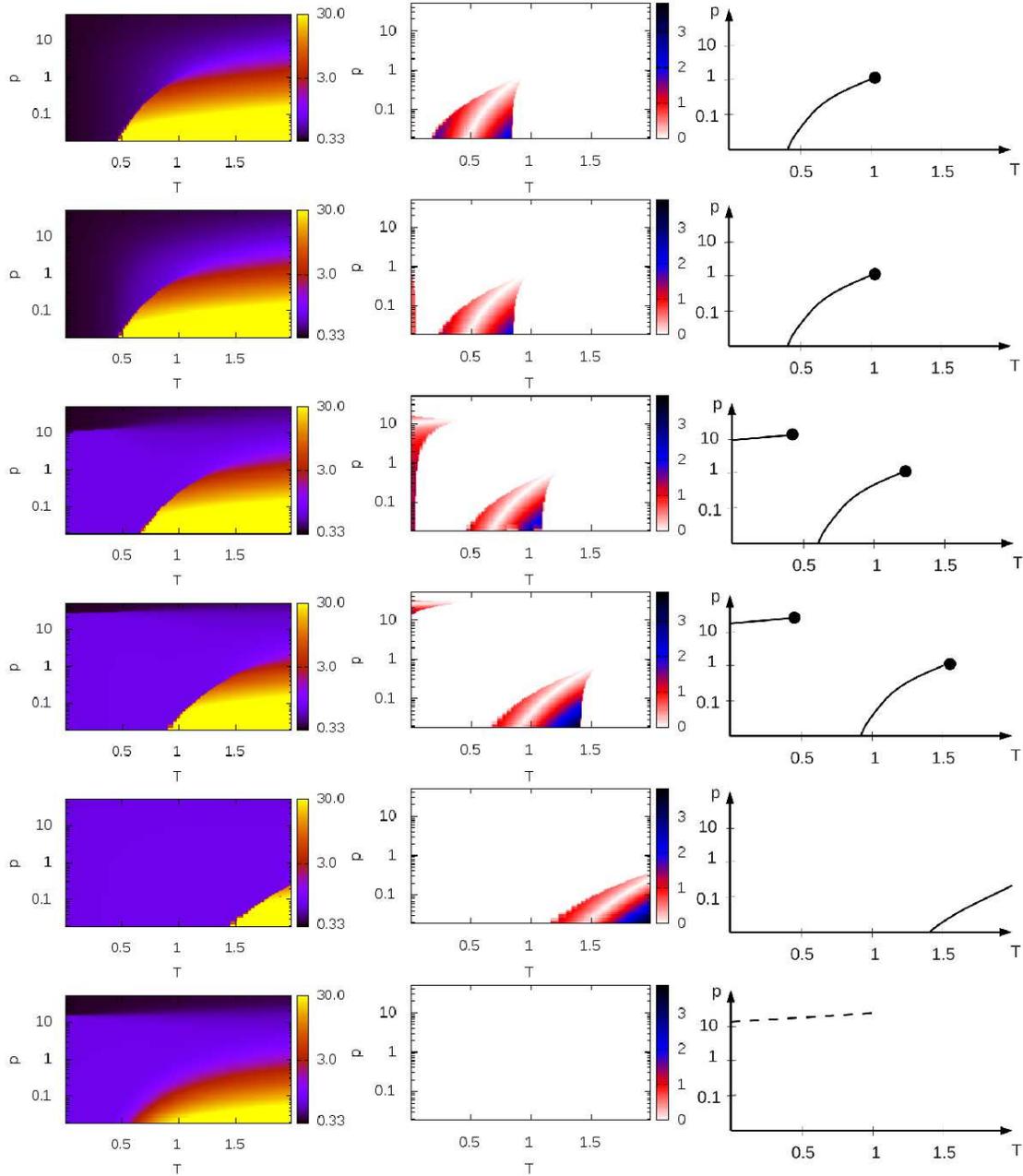}
\caption{Density profiles (left), coexistence regions (center), and phase diagrams (right) for $\epsilon = 0,-1,-3,-5,-10,-\infty$ (from top to bottom). The color code in the density profiles (left column) shows the specific volume $v$ for each state point, while in the coexistence region (central column) indicates the absolute value of the difference $|\Delta g|$ in the Gibbs energy between the coexisting phases. In the bottom line, which represents the pure Ben-Naim model, unscaled parameters $p$ and $T$ are shown, and the value $\epsilon=-5$ was used. Since the Ben-Naim model shows no phase transition, the location of steepest change in the specific volume in the ``phase diagram'' is indicated by a dotted line.}
\label{fig:fig1}
\end{figure*}

\begin{figure*}
 \subfigure[]{\label{fig:1_7_volume_eps1_n}\includegraphics[width=0.45\textwidth]{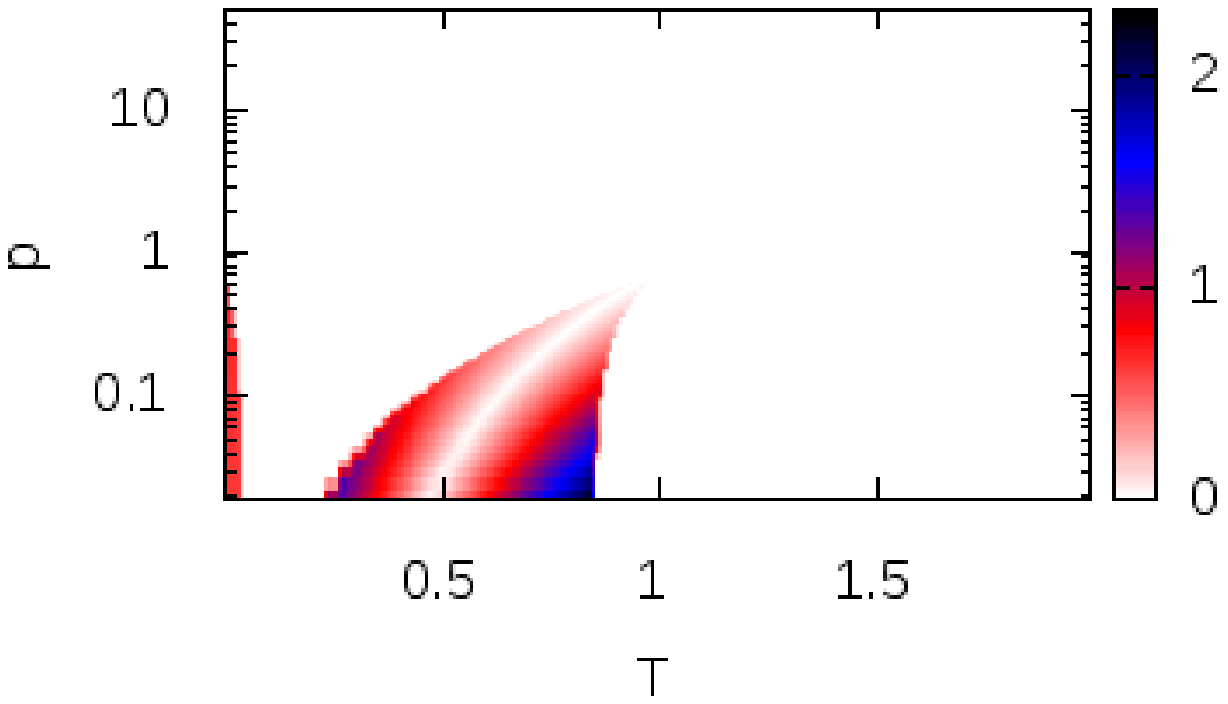}}
   \subfigure[]{\label{fig:1_7_volume_eps1p25}\includegraphics[ width=0.45\textwidth]{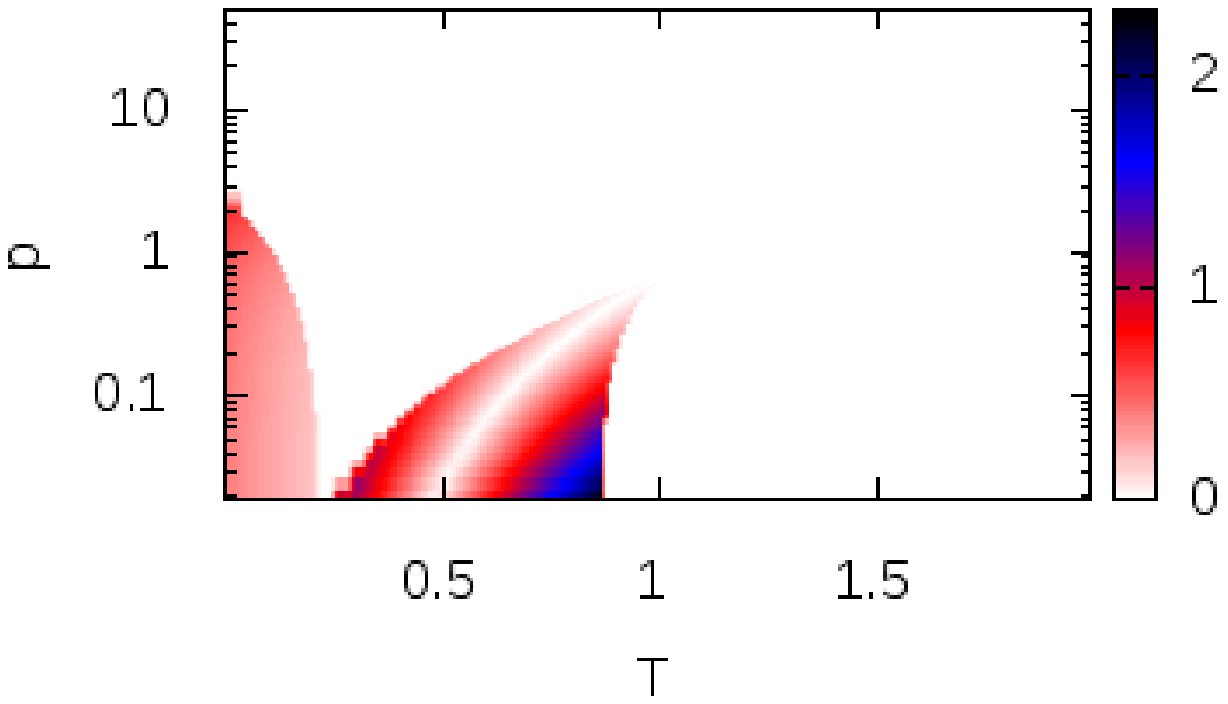}}\\
 \subfigure[]{\label{fig:1_7_volume_eps1p5}\includegraphics[width=0.45\textwidth]{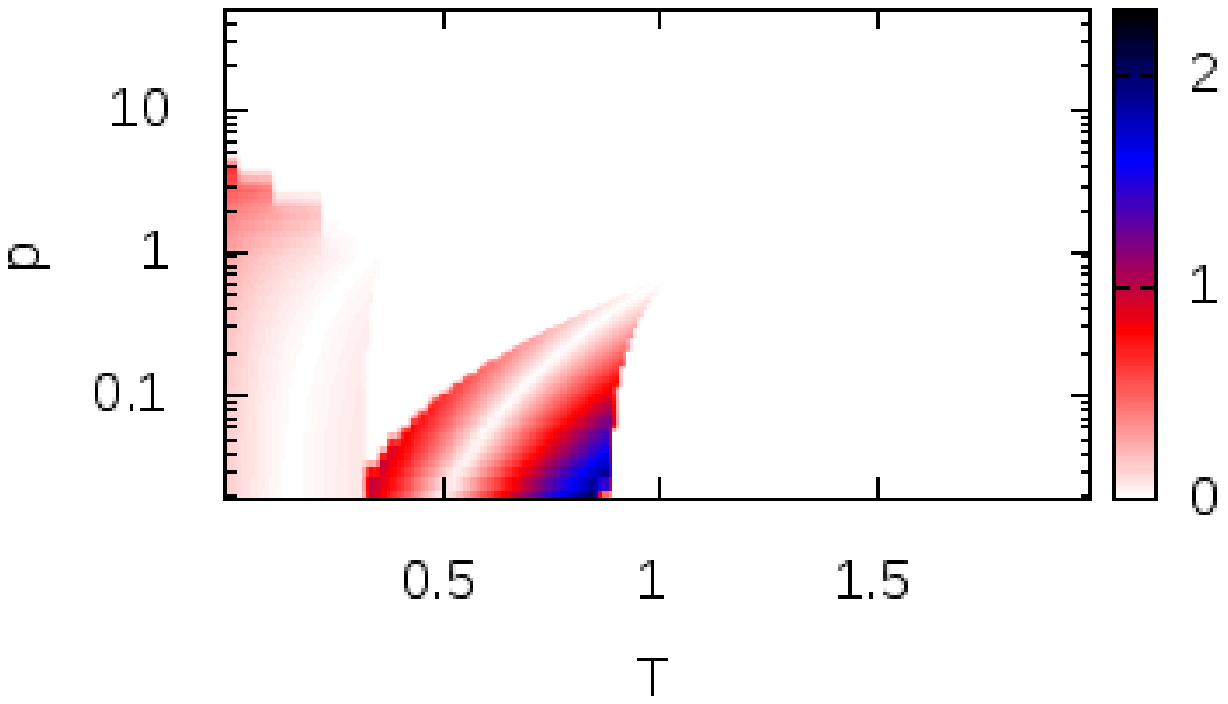}}
   \subfigure[]{\label{fig:1_7_volume_eps1p75}\includegraphics[ width=0.45\textwidth]{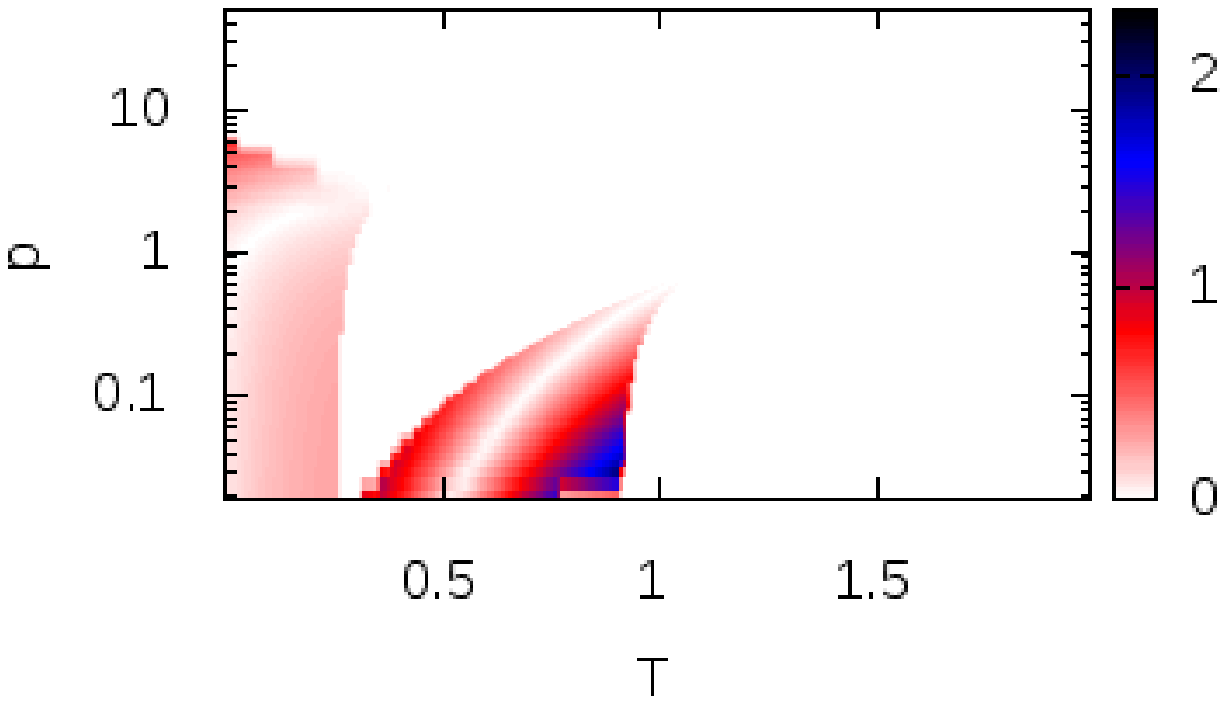}}\\
  \caption{Coexistence regions for the combined Ben-Naim--van der Waals potential for $-\epsilon = 1, 1.25, 1.5, 1.75$ (from (a) to (d)).}
    \label{fig:1_7_volume_eps1px}
\end{figure*}

\begin{figure}
\includegraphics[width=0.45\textwidth]{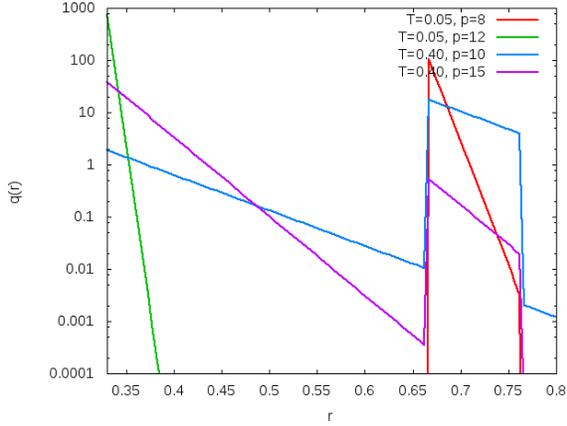}
  \caption{Probability $q(r)$ for a distance $r$ between two neighbouring particles for $\epsilon=-3.0$ and different values of $T$ and $p$.}
    \label{fig:5_1_qr_eps}
\end{figure}

\begin{figure}
 \subfigure[]
 {\label{fig:2_qis_eps3_q11}\includegraphics[width=0.3\textwidth]{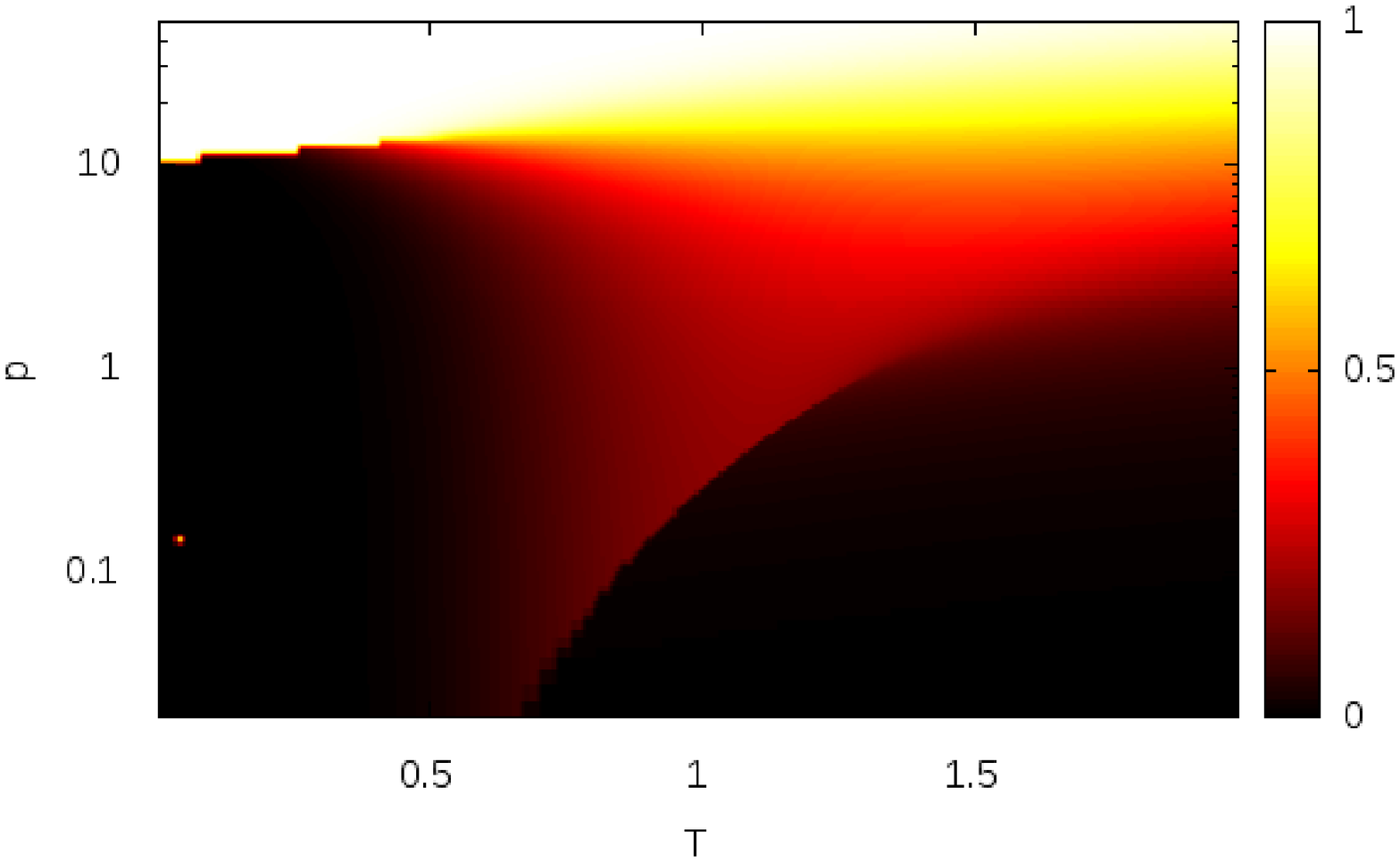}}
  \subfigure[]
 {\label{fig:2_qis_eps3_q12}\includegraphics[width=0.3\textwidth]{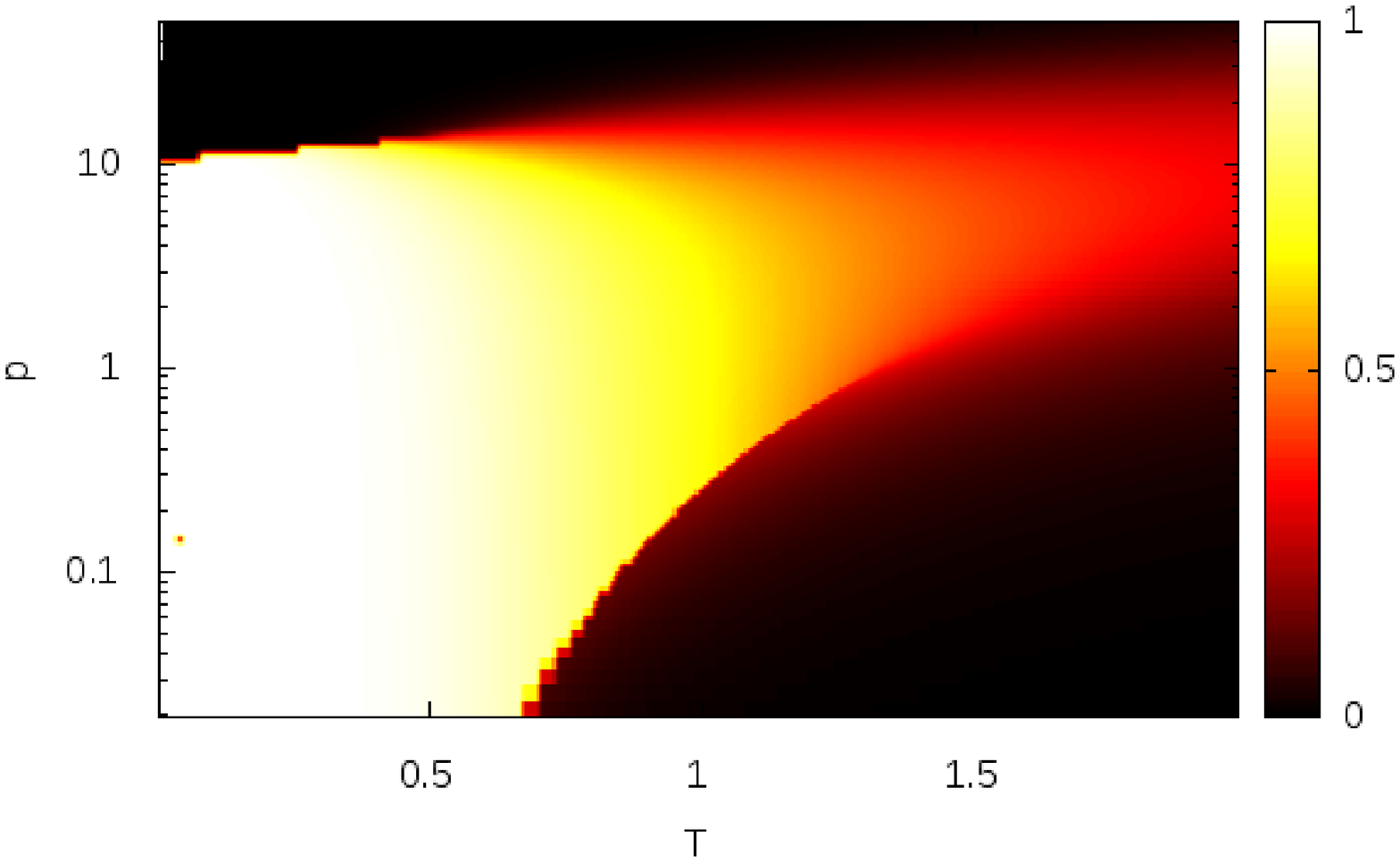}}
  \subfigure[]
 {\label{fig:2_qis_eps3_q13}\includegraphics[width=0.3\textwidth]{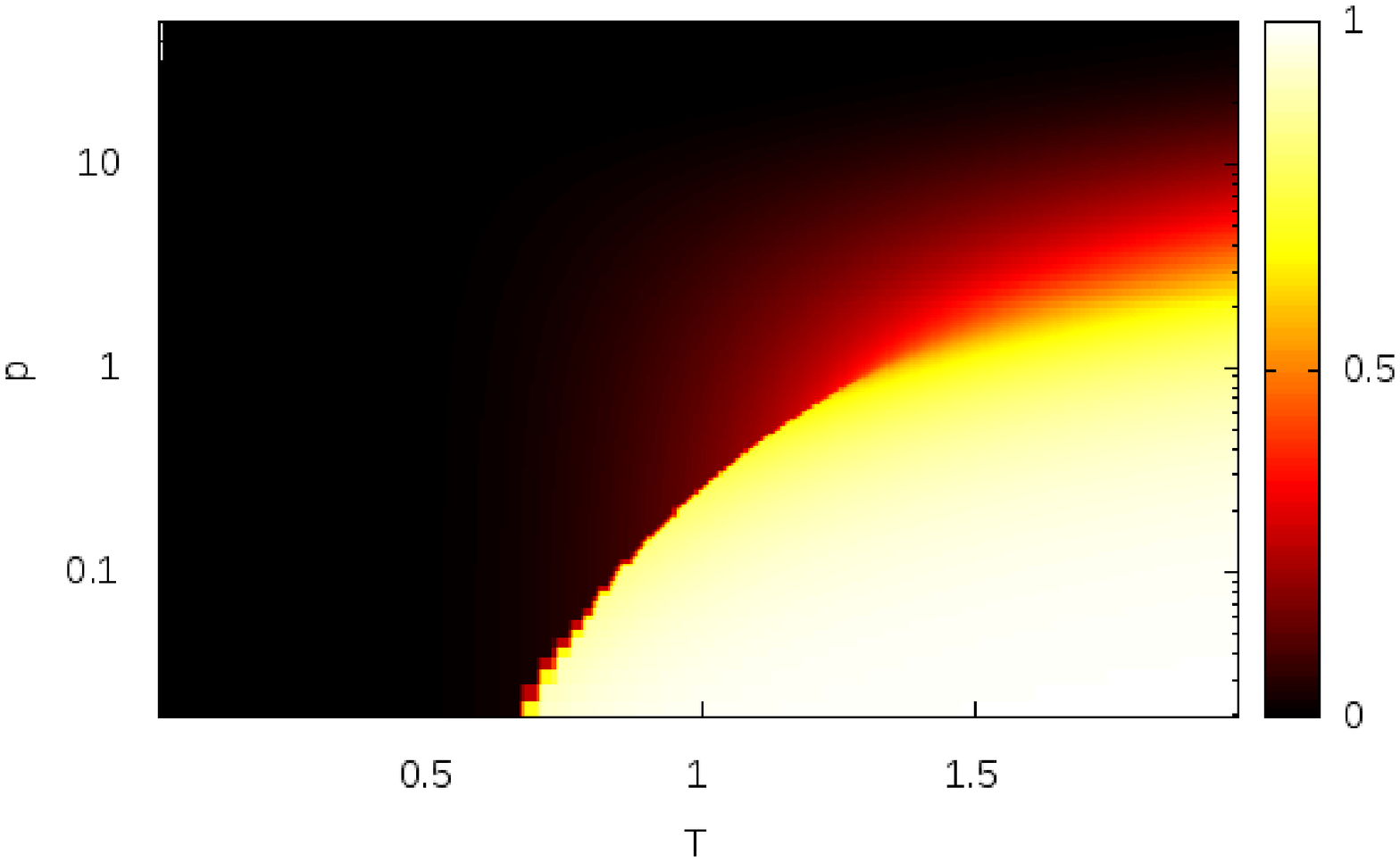}}
  \caption{Probability $q(r)$ for $\epsilon=-3.0$ and for a distance $r \in [\sigma_1, \sigma_2]$ (a), $r \in [\sigma_2, \sigma_2+\delta]$ (b), $r \in [\sigma_2 + \delta, \infty]$ (c).}
    \label{fig:2_qis_eps3}
\end{figure}

\begin{figure}[h!]
 \subfigure[]{\label{fig:1_4_volume_eps3_d001}\includegraphics[width=0.45\textwidth]{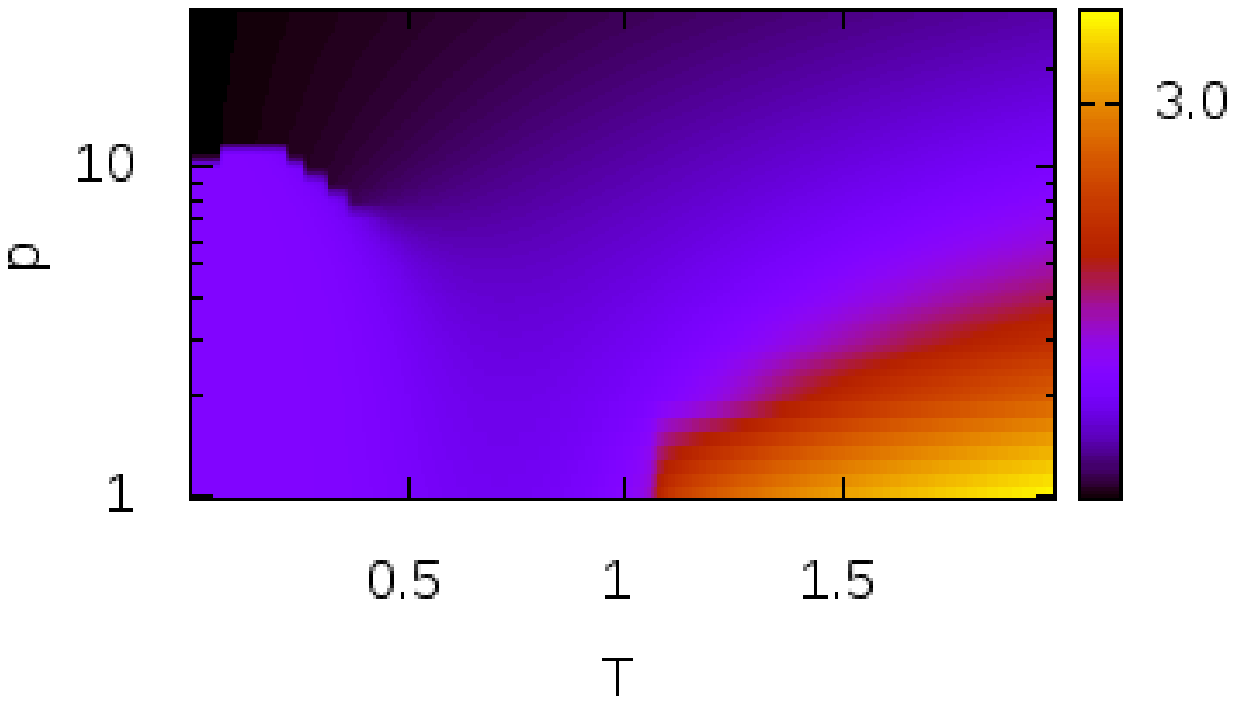}}
   \subfigure[]{\label{fig:1_4_volume_eps3_d0002}\includegraphics[ width=0.45\textwidth]{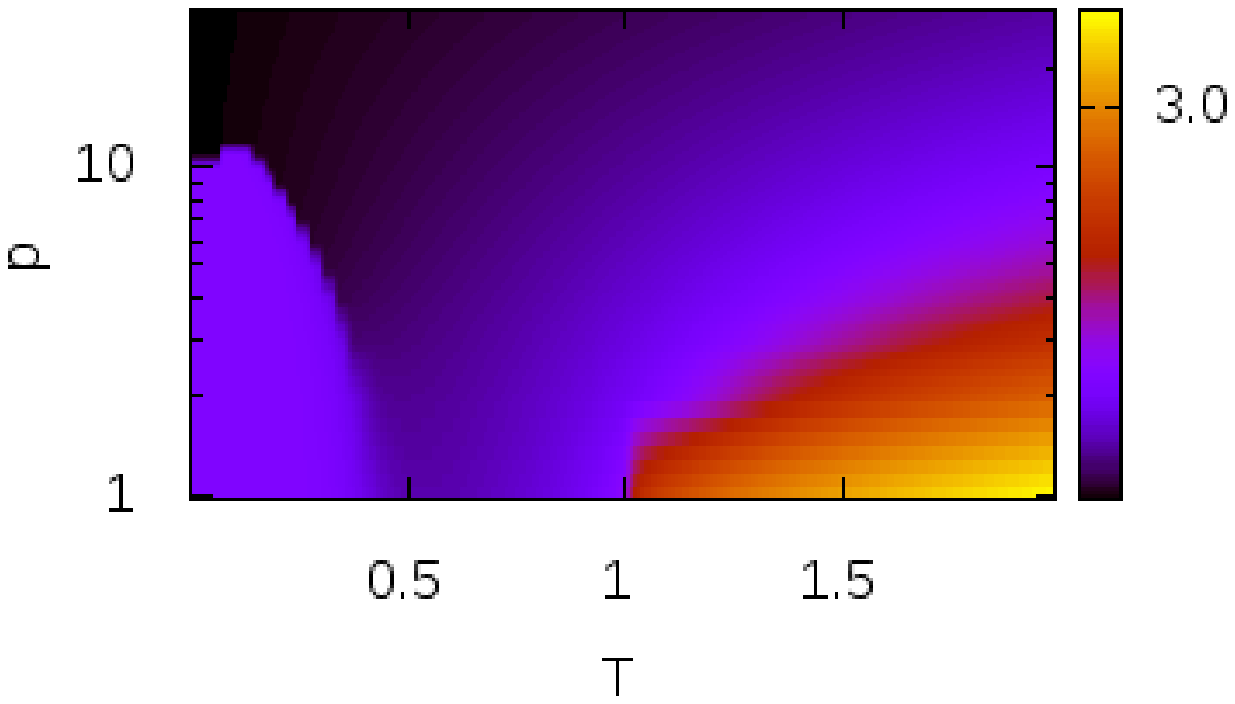}}\\
 \caption{Density profiles for $\epsilon=-3.0$ and (a) $\delta=0.01$  and (b) $\delta=0.002$.}
 \label{fig:1_4_volume_eps3_d}
\end{figure}

Figure \ref{fig:fig1} shows the density profile, the coexistence regions of two phases, and the phase diagram in the $p$-$T$ plane for different values of $\epsilon$. 
The bottom row shows the pure Ben-Naim model with $\sigma_1=1/3$, $\sigma_2=2/3$, $\delta=1/10$ and $\epsilon=-5$. For small pressures and temperatures, the specific volume is $ v\approx 0.67$, implying that most nearest-neighbor pairs are hydrogen bonded. At high pressure, the specific volume decreases to $ v\approx 0.34$, where the nearest-neighbor distances are close to the hard core diameter. At $T=0$, this transition is a real phase transition and occurs at $p=-\epsilon/(\sigma_2-\sigma_1)=15$.\\
The top row of Figure 1 shows the pure van der Waals model (where $\epsilon=0$). There is a first order phase transition (VdW-transition) between a phase with a small volume (corresponding to a high density fluid phase) and a phase with a higher volume corresponding to a low density gaseous phase. As temperature decreases or pressure increases, the density of the liquid phase increases continuously, however without displaying a region of particularly steep increase, as is the case in the Ben-Naim model.
The rows between the first and last show the changes in the system as $|\epsilon|$ ist increased. The first change that can be seen is the appearance of a metastable phase at very low temperatures.
This phase has a higher density than the stable phase, which is influenced by the presence of the potential well. With increasing $|\epsilon|$, the region where the metastable phase exists becomes larger (see Figure \ref{fig:1_7_volume_eps1px} for a zoom into the $|\epsilon|$ interval between $-1$ and $-2$), and at  $\epsilon \approx -1.5$ the metastable phase becomes stable for the first time.
This first occurrence of a phase transition between the two liquids can best be understood by considering the case $T=0$: 
At $T=0$ and going back to the original units, the Gibbs energy per particle in the high density phase is $g_h = p\sigma_1- a /{\sigma_1}$, and it is $g_l = \epsilon + p\sigma_2- a/{\sigma_2}$ in the low density phase. This corresponds to $g_h = {p\sigma_1}/{8}-{27}/{8\sigma_1}$ and $g_l = \epsilon + {p\sigma_2}/{8}-{27}/{8\sigma_2}$ in reduced variables. Since $p\sigma_2 > p\sigma_1$, the Gibbs energy of the high density phase can become smaller than that of the low-density phase only when 
 \begin{align}
-\epsilon  &\geq \frac{27}{8} \left( \frac 1 {\sigma_1} - \frac 1 {\sigma_2} \right) = 1.6875 \, 
 \label{eq:epsestimate}
 \end{align}
This is why the second phase transition occurs only for sufficiently large absolute values of $\epsilon$ and extends to $T =0$ only if $\epsilon < -1.6875$.
When $\epsilon$ increases further, the second transition is shifted towards higher pressures, and for $\epsilon=-10$ the transition is already out of the shown parameter range. 
The first (VdW-) transition is shifted towards higher temperatures with increasing $\epsilon$.
In order to better visualize the properties of the system on both sides of the phase transitions, we show in Figures \ref{fig:5_1_qr_eps} and \ref{fig:2_qis_eps3} the equilibrium distance distribution of particles for the value $\epsilon = -3$, where both phase transitions are pronounced. Figure    \ref{fig:5_1_qr_eps} shows the distribution $q(r)$ (see Equation  \eqref{eq:qofr})  for different temperatures and pressures corresponding to states below and above the LDL-HDL-phase transition. The distribution broadens with increasing temperature, and the probabilities for two particles to have the H-binding distance or the smaller distance, respectively, become more similar.
Figure \ref{fig:2_qis_eps3} shows the proportion of nearest neighbor pairs in each of the three distance intervals as a function of $T$ and $p$. 
One can clearly see that for high temperatures and low pressures, most particles have a distance larger than $\sigma_2+\delta$, while for low temperature and pressure most particles sit in the potential well, and for low $T$ and high $p$ most particles have the minimum distance $\sigma_1$. 
We end this section by discussing in more detail the LDL-HDL transition of our model. The slope of the LDL-HDL phase transition should have a negative sign for water, since the low-density phase is associated with a higher degree of order, while the high-density phase is less ordered. This is different from most other materials and is rather difficult to obtain in a simplified model, because in two or more dimensions a larger average particle distance corresponds to more available states. This is not true, however, for our one-dimensional model.
The slope of the phase transition can easily be changed in this model by adjusting the width of the well, $\delta$, as is shown in Figure \ref{fig:1_4_volume_eps3_d}. The slope of the phase transition is negative for sufficiently small ${\delta}$.

 \subsection{Anomalies}

\begin{figure}[h!]
 \subfigure[]{\label{fig:1_4_volume_eps3_p064} \includegraphics[width=0.45\textwidth]{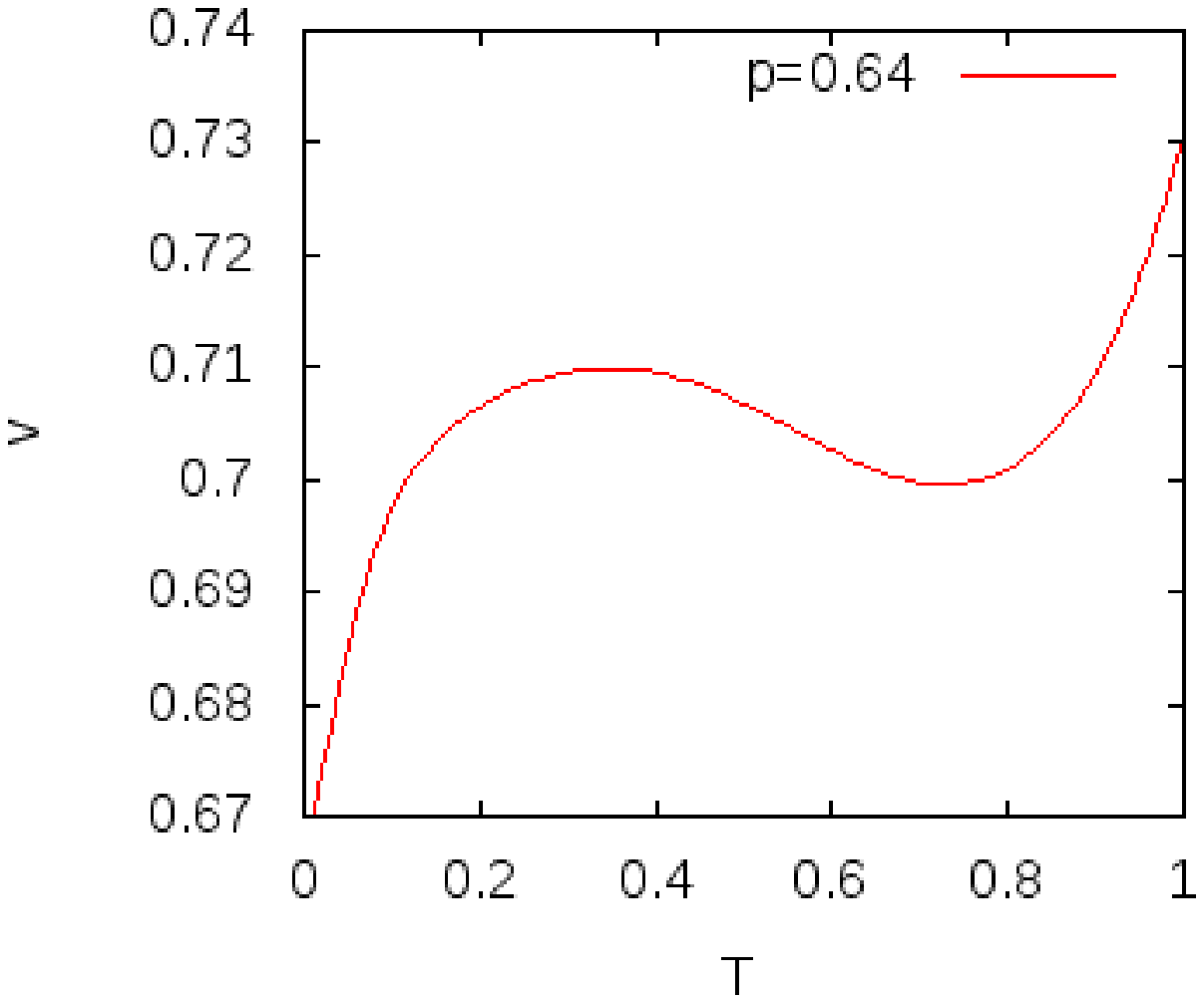}}
   \subfigure[]{\label{fig:6_1_ap_eps3_reg}\includegraphics[ width=0.45\textwidth]{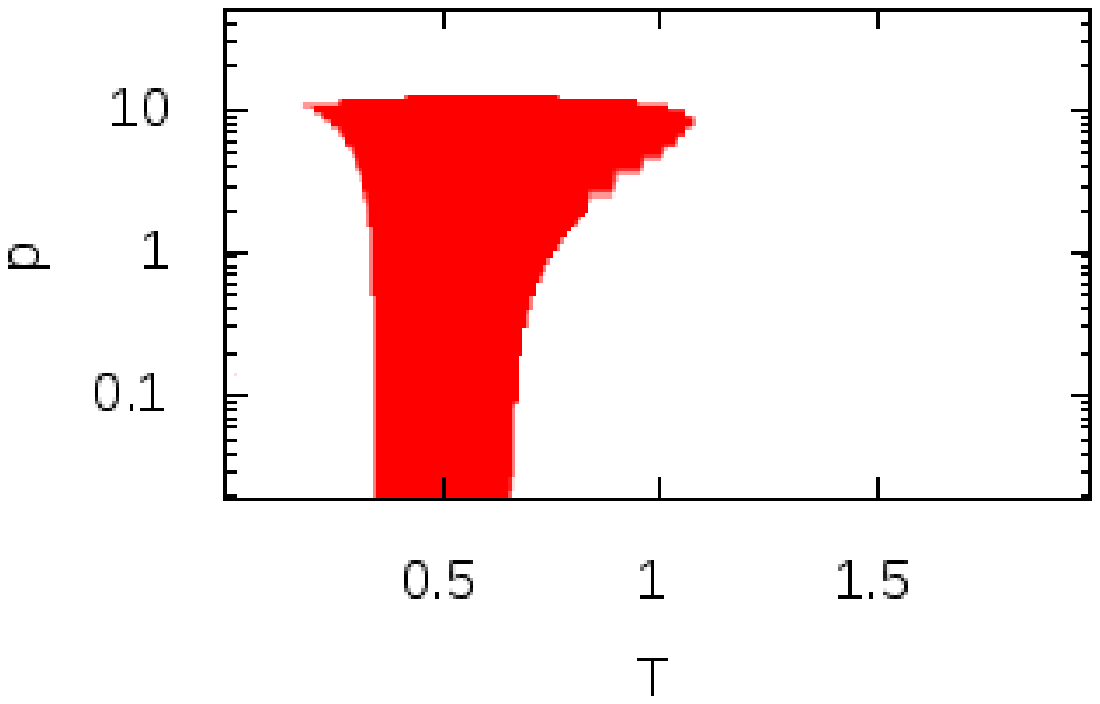}}
\caption{Density anomaly for $\epsilon=-3.0$. Figure (a) shows $v(T) $ exemplarily for $p=0.64$ and (b) shows the region where the coefficient of thermal expansion $\alpha_p = \frac{1}{v} \frac{\partial v}{\partial T}|_p$ is negative.}
 \label{fig:densanomaly}
\end{figure}

\begin{figure*}
   \subfigure[]{\label{fig:3_1_kt_eps3}\includegraphics[width=0.45\textwidth]{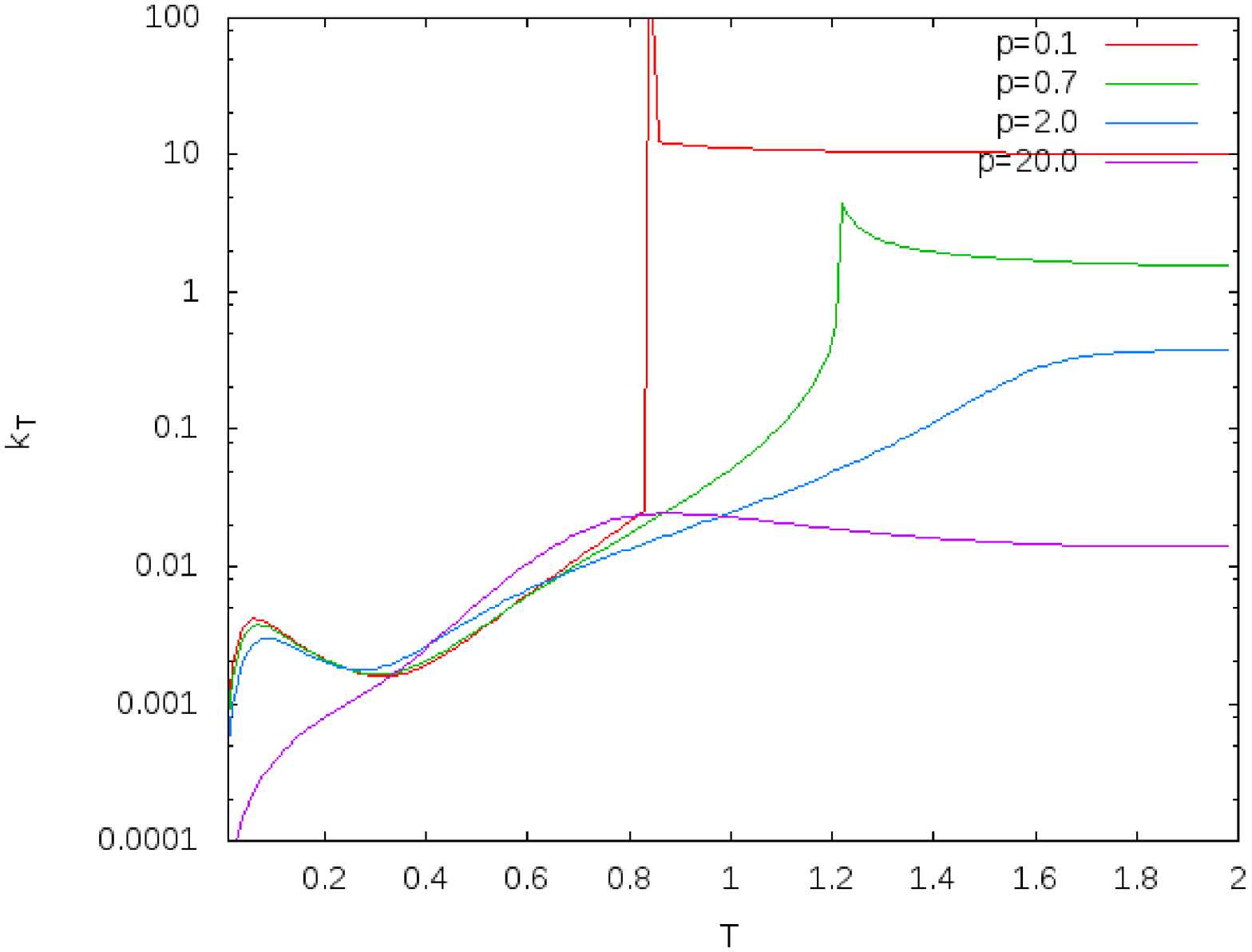}}
 \subfigure[]{\label{fig:4_1_cp_eps3}\includegraphics[width=0.45\textwidth]{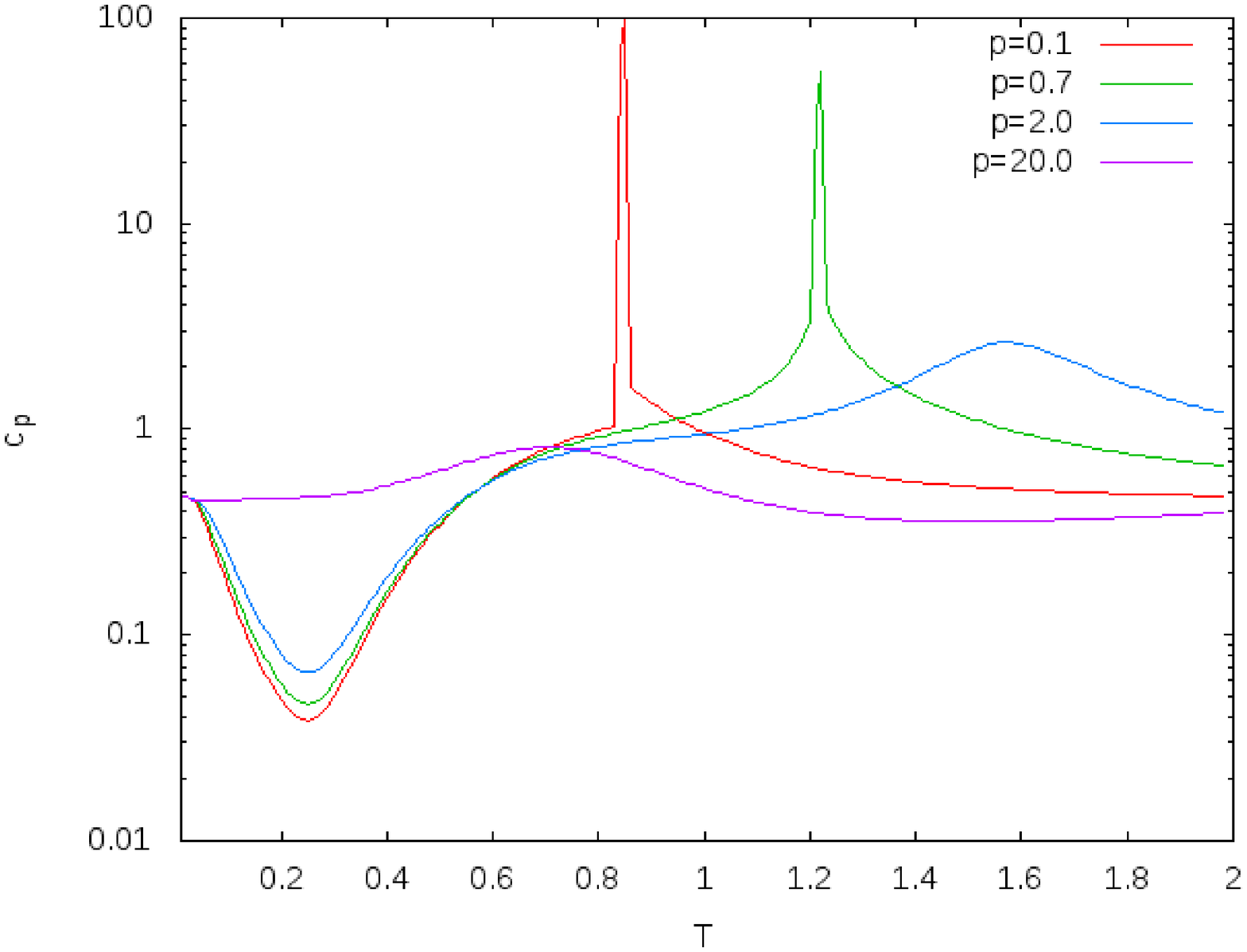}}\\
   \subfigure[]{\label{fig:3_1_kt_eps3_p}\includegraphics[width=0.45\textwidth]{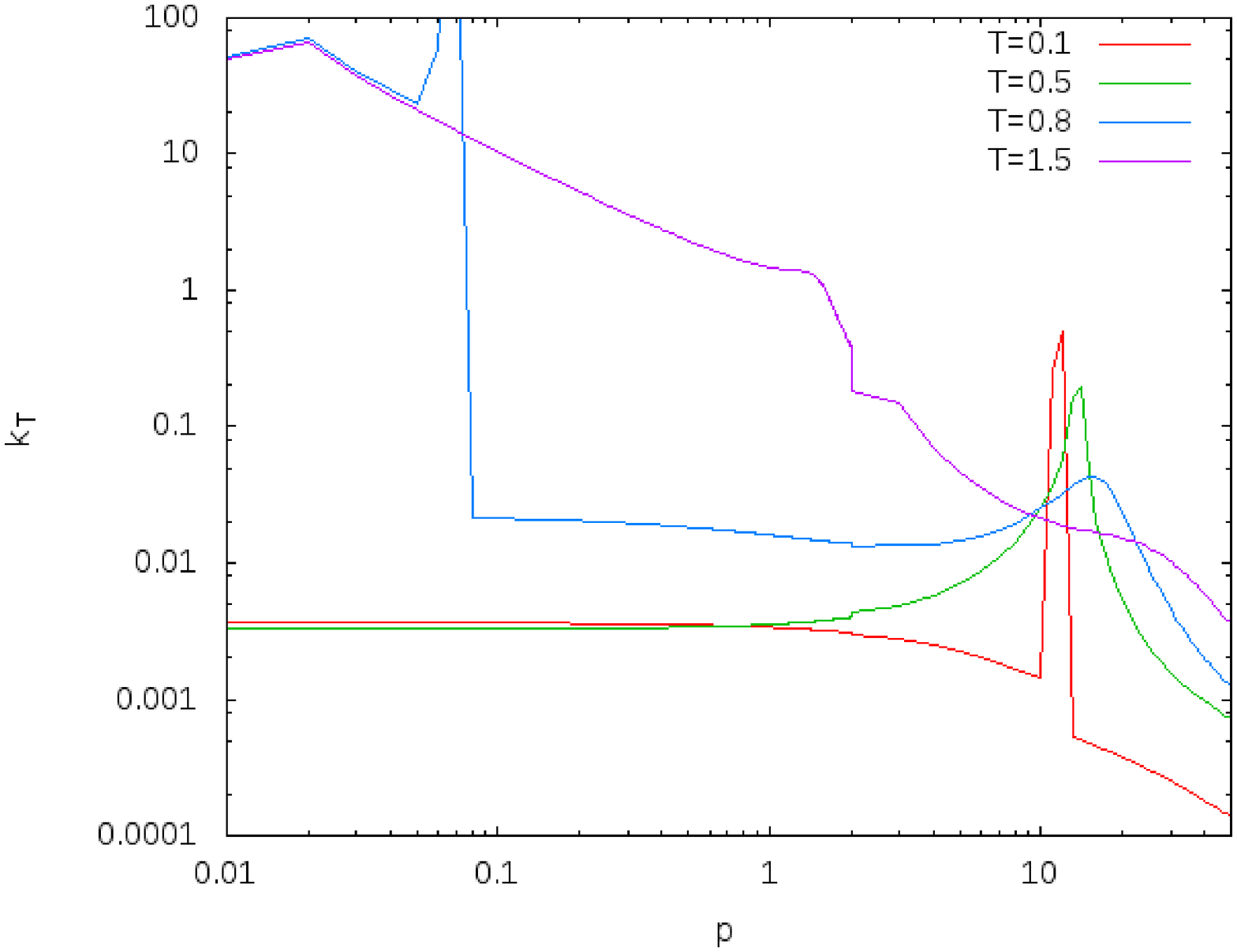}}
 \subfigure[]{\label{fig:4_1_cp_eps3_p}\includegraphics[ width=0.45\textwidth]{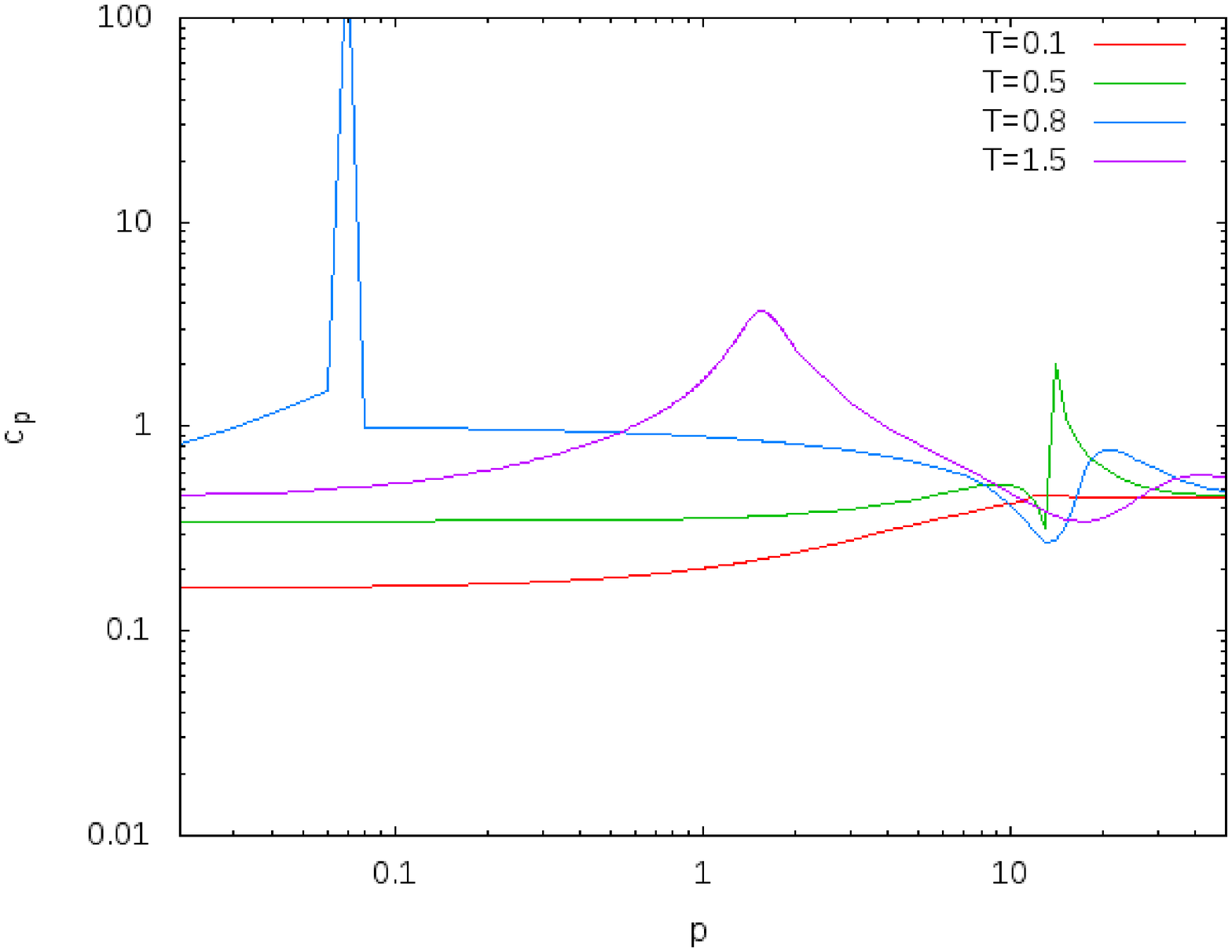}}\\
  \subfigure[]{\label{fig:3_2_kt_eps3_reg_sep}\includegraphics[width=0.45\textwidth]{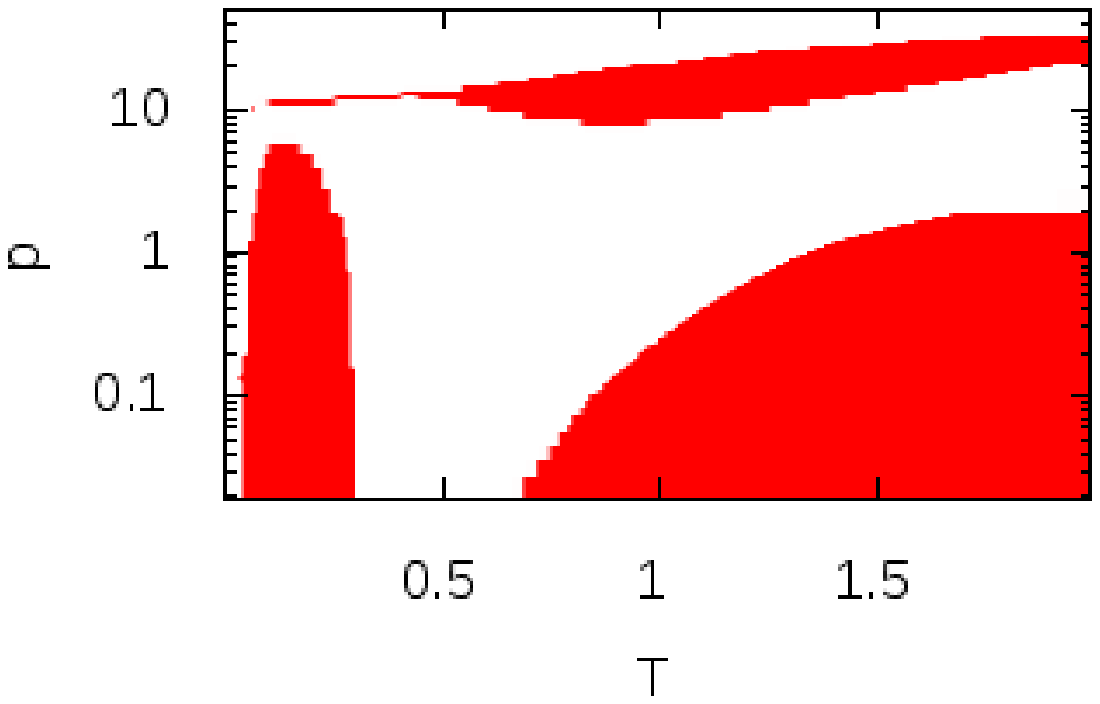}}
   \subfigure[]{\label{fig:4_2_cp_eps3_reg_sep}\includegraphics[ width=0.45\textwidth]{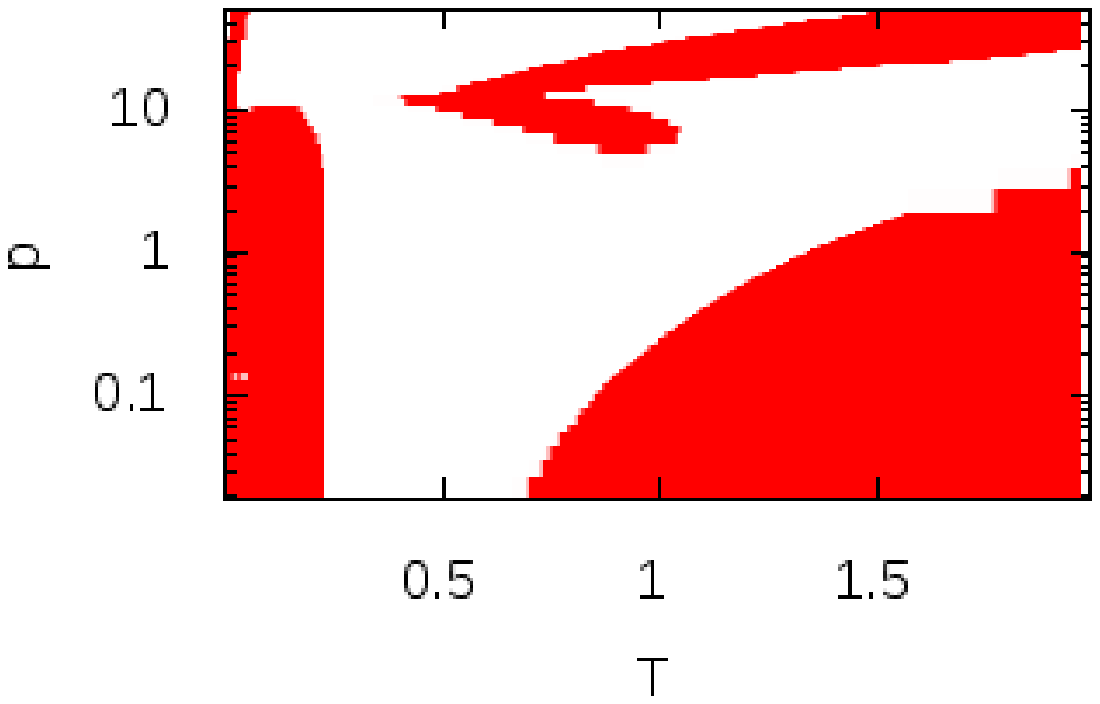}}\\
  \caption{Left column: Isothermal compressibility $\kappa_T$ as a function of $T$ (a), as a function of $p$ (c,) and regions where $\kappa_T$ as a function of $T$ has a negative slope (e). Right column: Isobaric heat capacity $c_p$ as a function of $T$ (b) and as a function of $p$ (d), and regions where $c_p$ as a function of $T$ has a negative slope (f). All plots are for a potential with $\epsilon=-3.0$.}
    \label{fig:3_4_kt_cp}
\end{figure*}

The combination of a hard core and a potential well at a larger distance leads to various anomalies similar to those of water, as was shown by Ben-Naim. The number of anomalies of water is large, see for example \url{http://www.btinternet.com/~martin.chaplin/anmlies.html}, and not all of them can occur in a simple model.
In the following, we investigate  three different anomalies in our model, which  are also present in the original Ben-Naim Model. We choose again $\epsilon=-3.0$, since for this value both phase transitions are well visible.
The density anomaly is shown in figure \ref{fig:densanomaly}. The specific volume has  a region of negative slope and a local minimum, where density is maximum. The region of negative slope, where the thermal expansion coefficient 
\begin{align}
\alpha_p = \frac{1}{v} \frac{\partial v}{\partial T}|_p
\label{eq:alphap}
\end{align}
is negative, becomes broader with increasing $p$, due to the closeness of the second phase transition.  The right boundary of the red area corresponds to a local density maximum, while the left boundary corresponds to a local density minimum (as function of $T$). With increasing temperature, more and more particles move out of the potential well and preferentially to its left side (see also figure \ref{fig:2_qis_eps3}). Beyond the red region, more and more particles move to the right side. Above a pressure of $p \approx 10$, the density anomaly vanishes. This behaviour is also known from real water, which does not show anomalies at high pressures \cite{bridgman12}. 
The isothermal compressibility and the isobaric heat capacity are shown  in Figure \ref{fig:3_4_kt_cp}. For a simple hard-rod system, one would expect that $\kappa_T (T)$ is a monotonically increasing function, while $\kappa_T (p)$ is a monotonically decreasing function. As to $c_p $, this would be a constant for all pressures and temperatures in a hard rod system \cite{bennaim08I}.
In our model, due to the presence of the potential well, both quantities show as a function of $T$ a minimum at low pressure, which vanishes at high pressure, as for the density anomaly. 
The singularities in the curves $\kappa_T(T)$ and $c_p(T)$ for low pressure correspond to the van der Waals transition, they are shifted to higher temperature for increasing pressure and turned into maxima above the critical point (see figure \ref{fig:3_1_kt_eps3} and \ref{fig:4_1_cp_eps3}). 
The maximum for high pressure ($p=20$) at $T\approx 0.8$, which is visible in both quantities, is due to the closeness of the Ben-Naim transition. 

The maxima in $\kappa_T (p)$ and $c_p(p)$ (see figure \ref{fig:3_1_kt_eps3_p} and \ref{fig:4_1_cp_eps3_p}) are also due to the Ben-Naim transition, as can be seen by comparison with the phase diagram. 
For the high temperature $T=1.5$, the van der Waals transition is crossed, leading to singularities at $p\approx 2$.

The features corresponding to the Ben-Naim transition were observed similarly in the pure Ben-Naim model \cite{bennaim08I}.
For instance, the minima of $\kappa_T (T)$ and $c_p (T)$ for small temperatures are also present without the van der Waals term. 
The maximum of $\kappa_T (p)$ at the BN-transition that was observed in the pure model is still present above the critical point at the so-called Widom line, but shows a discontinuity in the region of the phase transition as is expected for a real phase transition.
The modulation of $c_p (p)$ that was observed for the pure model \cite{bennaim08I} around the BN-transition is not visible in our data, but this may be due to the lower resolution we employ for the pressure. The overall behaviour of $c_p (p)$ shows also the signatures of both phase transitions.
The signatures of the van der Waals transition (discontinuities and maxima in the response functions) were, of course, not seen in the pure model.
To illustrate the behaviour of the response functions $\kappa_T$ and $c_p$ more globally, we show in figures \ref{fig:3_2_kt_eps3_reg_sep} and \ref{fig:4_2_cp_eps3_reg_sep} the regions where these variables have a negative slope as functions of the temperature $T$.
Just as for the density anomaly, negative slopes ${\partial \kappa_T}/{\partial T}<0$ and ${\partial c_p}/{\partial T}<0 $ are considered as 'anomalous' behaviour. 
One can see that both functions have three regions of negative slope: The region at small $T$ and small $p$, where the potential well traps particles that would otherwise be to its left; the region close to the HDL-LDL phase transition and around the Widom line; and third the region above the van der Waals liquid-gas transition. The negative slopes next to the phase transitions are expected since both $\kappa_T (T)$ and $c_p(T)$ have a maximum at the phase transitions.
A classification of parameter regions of anomalous properties of real water was made by Errington and Debenedetti \cite{1_2001_debenedetti}, who also investigated the interdependence of structural, dynamic and thermodynamic anomalies. 
The region of thermodynamic anomalies, where the density increases upon increasing temperature, spreads over a smaller density interval when temperature increases. Correspondingly, we find in our model that with increasing temperature, the region of pressure where an anomaly occurs gets narrower.
The region of anomalous density behaviour of three-dimensional particles interacting via isotropic core-softend potentials was investigated by de Oliveira and coworkers \cite{oliveira2006}.       
The line in the $p$-$T$ diagram corresponding to the temperature of maximum density has a similar shape as the curve we find for temperatures above the BN critical point.
The anomalies of our simple one-dimensional model resemble thus in several respects those of real water and water models although not all properties of water anomalies are matched. We want to emphasize two features that seem to be particularly interesting with respect to real water.
First, the region of anomaly of $\kappa_T(T)$ and $c_p(T)$ is restricted to low pressures and the system behaves normally at high pressures, which is similar in real water \cite{bridgman12}.
Second, the anomalies are not independent of the phase transitions. We observed that the coefficient of thermal expansion $\alpha_p (T)$ is strictly positive as long as $\epsilon$ is close enough to zero such that there is no BN-transition. 
However, when a phase transition occurs (as for $\epsilon =-3.0$, for example), the density anomaly is also seen in a temperature region above the critical point of this transition (compare figure \ref{fig:fig1} and \ref{fig:6_1_ap_eps3_reg}). This supports the hypothesis that anomalies of water may be related to phase transitions at a lower temperature that is not observable due to spontaneous crystallization, but gives rise to signatures in form of anomalies at ambient conditions.

\section{Discussion and Conclusion}

We have introduced a simple one-dimensional model that combines two existing models and shows several properties of real water. The model has the following two features: 
1.) A short-range potential that introduces two length scales and correlates low binding energy with a low density and thus with an open structure \cite{bennaim08I} and
2.) a long-range interaction that lowers the energy when the total volume is smaller.
These are two properties that real water possesses: The short range potential leads to the formation of hydrogen bonds that correlate low binding energy with an open structure. Additionally, there is a long-range interaction from electrostatic interactions and van der Waals forces that encourages high density. 
In contrast to the pure Ben-Naim model, which has only short-range interactions and therefore no phase transition, our model has two first-order phase transitions, a liquid-gas transition and a HDL-LDL transition. An important parameter of the model is the ratio of the two energy scales that are given by the depth of the potential well and by the strength of the mean-field attraction. Only when this ratio $\epsilon$ is high enough does the second  phase transition occur. With increasing $\epsilon$, the two phase transitions move to higher $T$ and higher $p$, respectively.
The relative importance and balance between the long-range van der Waals force and a short range hydrogen bonding is presently discussed for ice \cite{santra2011}. 
Also in a one-dimensional lattice model the balance between the mean field attraction and the short-range interaction potential had to be adjusted in order to obtain a phase diagram similar to water \cite{hoye2010}.
Of course, a one-dimensional model cannot be a complete and realistic description of real water, and it cannot make quantitative predictions. However, a simple model can help to gain insight into the principles underlying the special properties of water, and it may lead to more understanding than a more complex model. The partition function and implicit expressions for various thermodynamic quantities could be given analytically, and were evaluated numerically. The results obtained using simple models, which can be written down analytically, can be interpreted much easier  than those obtained with more complex models, which would require computer simulations \cite{dill2005}.
A generalization of our model to higher dimensions could be done in several ways, all of which have their drawbacks. In particular, the geometry must be considered. If the potential would be generalized to three dimensions by making it spherically symmetric, the number of states in the potential well would be much larger than at shorter distances, leading to entropic effects. For example, the slope of the HDL-LDL transition would almost certainly be positive because of the Clausius-Clapeyron relation.
In investigations of the two-scale Jagla-Potential in three dimensions, for example, a LDL-HDL-transition was found, but also with a positive slope in the $p$-$T$-plane \cite{xu2006}.

In order to compensate for these entropic effects, angle-dependent potentials would have to be introduced as for example in a very recent publication by the Stanley group \cite{tu2012}, where a Widom line with a negative slope was found for strong tetrahedral interactions.
This approach, however, is already very close to the detailed modelling of single molecules, where a tetrahedral geometry emerges from the presence of two hydrogen atoms arranged in a nearly tetrahedral angle with respect to oxygen, and thus provides not much more insight compared to molecular models. Also, it would be necessary to know whether the four sites where H-bonds can be formed are independent. This seems to be the case according to Predota et al. \cite{predota2003}.

The relation between the dimension of the model, anomalies, phase transitions and the interaction potential is discussed by Buldyrev et al. \cite{buldyrev2002}. In this article, a double-step-potential is investigated in one, two and three dimensions and also the difference between two- and three-dimensional models is discussed.
The authors find in their model systems that liquid anomalies and a liquid-liquid phase transition may occur independently and that a density anomaly in a low density phase is not seen when the LDL-HDL phase transition line has a positive slope.
We can not confirm these effects as general trends, as we observe the density maximum only when a LDL-HDL transition is present and it occurs in the LDL phase. However, we agree about the shape of the temperature of maximum density $T_\rho (p)$ and about the fact that this function has a maximum.

Models in more than one dimension do not require a long-range attraction in order to show phase transitions. Therefore, most higher-dimensional models have only short-range potentials, but all of them require two different length scales. Indeed, it has been discussed by various authors that a two-length-scale potential is a necessary ingredient in order to obtain thermodynamic anomalies similar to water \cite{oliveira08} and there is evidence that the hierarchy of anomalies is determined by the relation between the two length scales \cite{fomin2011}. The exact form of the two-length-scale potential seems not to be important for the occurrence of anomalies, since they occur also in a model with a repulsive step \cite{oliveira08} instead of an attractive well as in the present manuscript. In any case, the interaction potential has to be such that a higher density is correlated with a higher energy and a lower density with a lower energy, and our work confirms that this correlation is necessary for the presence of anomalies. 
However, water is a complex liquid and there is some evidence that simple principles for the interaction potential may not be the whole story.
Errington et al.\cite{errington2006} and later Yan et al.\cite{yan08_PRE78} suggested that the occurrence of anomalies can be related to and predicted by the excess entropy, implying that this quantity may be more relevant for the occurrence of anomalies than the shape of the interaction potential.
It has to be noted, however, that some assumptions of Errington's work do not hold in anomalous regions \cite{fomin2010}.
Another recent article on this topic states that a two-length-scale potential may not fully account for anomalies, but that energetic and entropic effects may be relevant as well \cite{salcedo2011}.

As a final note, we would like to point out that there exist other materials besides water that exhibit anomalies. Anomalies in silica are for example investigated by Shell et al. \cite{shell02}, and  in the work by Hoye and Lomba \cite{hoye2010} the comparison of water with other tetrahedral substances such as Si or Ge was made. In a publication by Angell et al. \cite{angell2000}, further tetrahedral liquids are mentioned. The simple model discussed in this paper, as well as other simple models, do thus not only help to understand water, but also other materials that have similar anomalies.

\section*{Acknowledgments}
This work was supported by DFG Grant Dr300/11.

\end{document}